\title{BRICS Astronomy and the United Nations Open Universe Initiative}
\author{
        Ulisses Barres de Almeida \\
        Brazilian Center for Physics Research (CBPF)\\
        Rua Dr. Xavier Sigaud 150, Rio de Janeiro 22290-180, \underline{Brazil}
            \and
       Paolo Giommi\\
       Italian Space Agency (ASI)\\
       Via del Politecnico snc, 00133 - Roma, 
       \underline{Italy}
            \and
       Andrew M.T. Pollock\\
       University of Sheffield, Hounsfield Road, Sheffield S3 7RH,
       \underline{England}
}
\date{\today}

\documentclass[12pt]{article}
\usepackage{graphicx}
\usepackage{hyperref}

\begin{document}
\maketitle

\textbf{This paper is being submitted to the Journal ''Annals of the Brazilian Academy of Sciences' as part of the Proceedings for the BRICS Astronomy Workshop  - BAWG 2019.}

\begin{abstract}
The almost universal availability of electronic connectivity, portable devices, and the web is bringing about a major revolution: information of all kinds is rapidly becoming accessible to everyone, transforming social, economic and cultural life practically everywhere in the world. Internet technologies represent an unprecedented and extraordinary two-way channel of communication between producers and users of data. "Open Universe" is an initiative proposed to the United Nations Committee on the Peaceful Uses of Outer Space (COPUOS) and currently in implementation under the leadership of the United Nations Office for Outer Space Affairs (UN-OOSA). Its primary objective is to stimulate a dramatic increase in the availability and usability of space science data, extending the potential of scientific discovery to new participants in all parts of the world. This paper describes the initiative in general, some of the activities carried out to demonstrate its feasibility, and its use in the context of the BRICS Astronomy Programme.
\end{abstract}

\section{Introduction}
In the field of space science almost all existing data sets have been produced through public funding, therefore they should be considered a public good and become openly available to anyone at a certain point in time. In particular, high-level calibrated data products, like images, spectra and similar products, should be available in a transparent form, that is usable by all. To ensure a fair scientific reward and to protect the intellectual property of the teams that conceive, design, build and operate the instruments that generate the data, this should happen according to clear rules.

The benefits of openness and transparency have been widely emphasised. They are so large and evident for both users and data providers that even scientific space data generated through private funds should aim at transparency.

Much has been done in recent years, especially in space astronomy, to offer open access, user-friendly platforms and services, demonstrating how natural is the evolution towards a more and more transparent and inclusive ecosystem of tools and services. However, despite the recent progress, there is still a considerable degree of unevenness in the services offered by providers of space science data.

Further efforts are necessary to consolidate, standardise and expand services, promoting a significant inspirational data-driven surge in training, education and discovery. Such a process, leading to a much larger level of availability of space science data, should be extended to non-scientific sectors of society.

To respond to these needs, at the fifty-ninth session of the United Nations Committee on the Peaceful Uses of Outer Space (COPUOS), the Government of Italy, working closely with the Italian Space Agency (ASI), proposed the “Open Universe Initiative”~\cite{proposal}. Brazil is today one of the leading members of "Open Universe", opening up the way for inclusion of the work of the BRICS Astronomy Working Group (BAWG, see below)  among the activities of the Initiative.

The far reaching vision of the Open Universe Initiative and the potentially global scope, which extends the benefits of space science to large sectors of the society, including emerging and developing Countries, call for a wide international cooperation under the auspices of the United Nations with activities fully integrated into the UN Space2030 agenda~\cite{space2030}.

In this paper we describe the main principles behind the Open Universe, whose costs are easily sustainable and certainly marginal with respect to the total investment that is made to produce astronomical and science data. In particular, we briefly address the potentially very large socio-economical returns of the initiative to the BRICS countries.

\paragraph{The BRICS Astronomy Working Group (BAWG)\protect\footnote{The BRICS Astronomy Working Group website can be found at the following link: \protect\url{https://www.bricsastronomy.org}.}} was created in 2015 with the mission of promoting cooperation between BRICS member countries in the field of astronomy, and to enable technologies through joint activities of government, universities, research institutions, and industry. The BAWG is composed of government officials and designated representatives, supported by the focal points on astronomy and experts from BRICS member countries. It provides a platform for BRICS member countries to engage on policy issues and other matters related to research, development and practice in astronomy, and to explore mechanisms for promoting BRICS cooperation in astronomy. The principal goal of the Group is to develop the astronomical sciences and generate new knowledge, training, in the process, human capital and developing new technologies and applications that can improve the public understanding of science. Such goals are greatly aligned with the principles and objectives of the Open Universe Initiative, in such a way that their cooperation for common goals is highly auspicious.

\section{The Open Universe Initiative}\label{UNOU}
“Open Universe” is an initiative under the auspices of COPUOS with the objective of stimulating a dramatic increase in the availability and usability of space science data, extending the potential of scientific discovery to new participants in all parts of the world and empowering global educational services.

Following the original proposal by Italy at the 59$^{th}$ session of COPUOS in 2016\protect\footnote{See COPUOS Document A/AC.105/2016/CRP.6}, and the adhesion of other supporting countries, such as Brazil, the initiative was welcomed as part of the Space 2030 Agenda~\cite{space2030}, to be carried out as an activity of UNOOSA towards the Sustainable Development Goals (SDGs). The initial phase of development of the Initiative was marked by a number of open discussions, including an expert meeting that took place at the Italian Space Agency (ASI) in April 2017\protect\footnote{\protect\url{openuniverse.asi.it/documents/ou\_documents.php}}, a series of presentations during COPUOS meetings or UN organized high-level fora, and a full workshop held at the UN premises in Vienna, in November 2017\protect\footnote{\protect\url{unoosa.org/oosa/en/ourwork/psa/schedule/2017/workshop\_italy\_openuniverse.html}}.

“Open Universe” will work to ensure that space science data will become gradually more openly available, easily discoverable, free of bureaucratic or administrative barriers, and usable by the widest possible community, from professional space scientists (several thousands of individuals) to citizen scientists (potentially of the order of millions), to the common citizens interested in space science (likely hundreds of millions).

\begin{figure}[th!]
  \includegraphics[angle=-90, width=\textwidth]{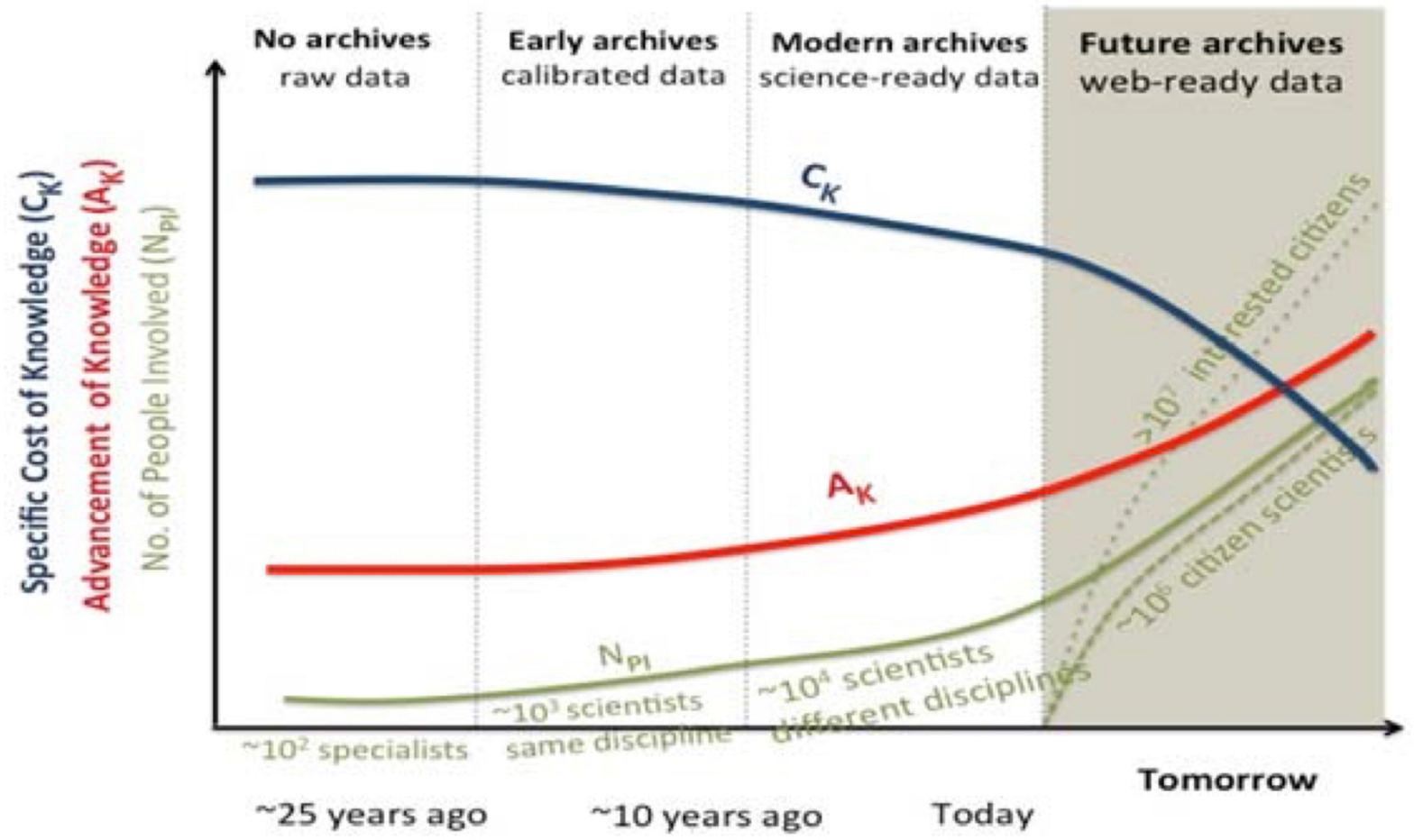}
  \centering
  \caption{Conceptual timeline of the evolution of data archives in Astronomy. Taken from the original Open Universe proposal A/AC.105/2016/CRP.6}
  \centering
   \label{Fig:timeline}
\end{figure}

The services delivered by existing space science data producers have significantly improved over time (see timeline provided in Figure \ref{Fig:timeline}), but are still largely heterogeneous, ranging from the basic support reserved to a restricted number of scientists, to open access web sites offering ”science-ready” data products, that is, high-level calibrated space science data that can be published without further analysis by professionals with suitable knowledge.

”Open Universe” will implement methods for improving the transparency and usability level of the data stored in current space science data archives, and urge the data producers to increase their present efforts so as to extend the usability of space science data to the non- professional community.

After a number of public discussions and international meetings, Open Universe is now on track for its formal foundation under the leadership of the UNOOSA, and in close cooperation with the Governments of Italy and Brazil, as well as in collaboration with other participating Countries and institutions.

\section{Socioeconomic benefit considerations}\label{benefits}
The Open Universe initiative aims at largely improving the level of transparency, and therefore accessibility and use of space science data both in the scientific community and in non-scientific sectors of the society. In both cases, socioeconomic benefits may be considerable: In the first case, in the form of an increased number of publications in scientific journals, especially by new actors to be admitted in the field from across the globe, and in the second case, in the form of e.g. better education and a larger discovery potential for a broader user-base.

A detailed cost-benefit analysis of the Open Universe initiative is certainly beyond the scope of this paper. In the following, we limit ourselves to a few considerations on the social and economic benefits of the initiative. 

\itemize{
\item Improve services to professional scientists
\item Facilitate cross-disciplinary, multi-experiment and data-intensive research
\item Improve international cooperation in space sciences
\item Increase the efficiency in the production of knowledge for the same data sets, thus reducing the specific cost of knowledge
\item Enable more citizen scientist-type activities
\item Promote scientific thinking and scientific culture in society
\item Inspire young people to scientific careers and stimulate capacity building for the industry 4.0
\item Improve the quality of STEM education
\item Extend the potential of scientific discovery to non-professional scientists, in principle to everyone interested in Space Science.\\
}

Many of the listed items have a potentially direct impact to the BRICS countries. Among them, promotion of international cooperation in space sciences is particularly relevant. In fact, according to the CAPES InCites Report 2017\protect\footnote{\protect\url{https://www.capes.gov.br/images/stories/download/diversos/17012018-CAPES-InCitesReport-Final.pdf}} Space Science is the research area with greatest citation impact in Brazil, a result which is strongly correlated with the high-degree of international collaboration in the field. In addition to that, although far from being the leader in number of publications, research conducted in collaboration with one or more of the BRICS countries have ranked as the most impactful (in terms of citations per paper) for Brazilian science across all disciplines\protect\footnote{The report in question thus recommends that "[...] a strategy of expanding strategic collaborations with the BRICS countries would yield greater dividends than collaboration with with North American and European countries, for instance." (InCites Report 2017, page 13)}. These factors very clearly highlight the potential in store for increasing BRICS cooperation in the field of space science, and the importance of open science in strengthening such multi-lateral cooperation.

Cost-benefit analysis of research infrastructures is a difficult matter and it is rarely attempted, mostly because of the unpredictability of future economic benefits of science. Nevertheless, it has been recently conducted on some large scientific projects, the most important one being the Large Hadron Collider (LHC), the world’s largest particle accelerator, showing that the benefits may be extremely large. For instance~\cite{Florio2016}, showed that the economic value of the two widely used open access softwares ROOT and GEANT4, developed and maintained within LHC activities, can be estimated to be 2.8 billion Euros.

\begin{figure}[th!]
  \centering
  \includegraphics[angle=-90, width=\textwidth]{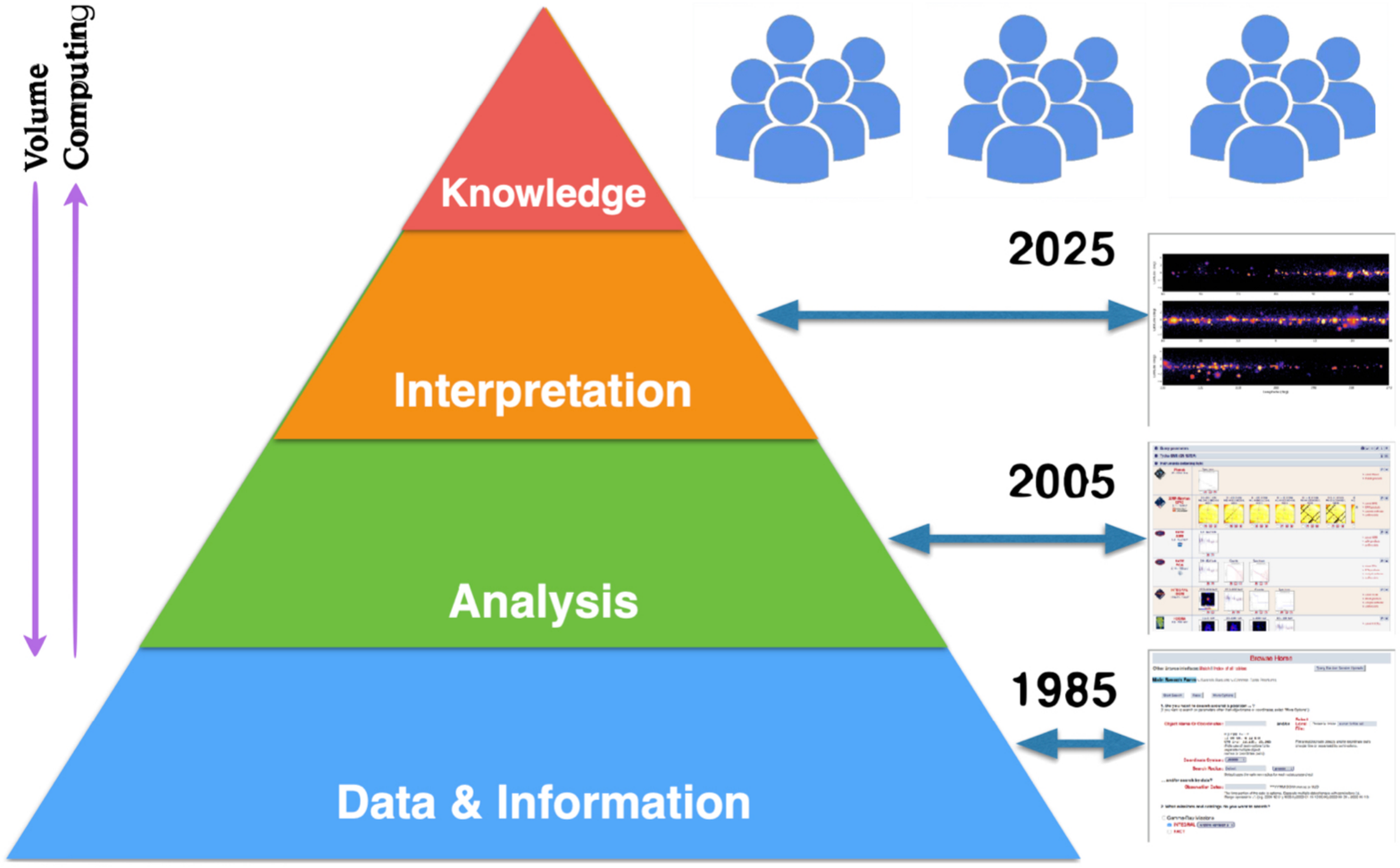}
   \caption{Diagram illustrating how the increase of computational processing of the data, to high-level science products, can induce increased accessibility to knowledge. Adapted from a figure by Roland Walter presented at the 2017 UN-Italy Workshop on the Open Universe Initiative.}
  \centering
   \label{Fig:pyramid}
\end{figure}

Expanding the use of space science data to large sectors of society may be achieved by lowering the barriers of usability of space science data to the level of every citizen, (that is offering web-ready or transparent data products) rather than following the more frequently used approach of educating a small fraction of citizens (usually students and researchers) to be able to perform complex analyses or data processing on low or intermediate level products like raw or un-calibrated data. Figure \ref{Fig:pyramid} illustrates the concept that, by increasing the level of computational processing of the data, to high-level science products, larger accessibility to knowledge is achieved.

The potential benefits are clearly very large and will depend on how the initiative (and other related activities, e.g. the Virtual Observatory) will evolve.

\section{The role of international cooperation}\label{cooperation}
International cooperation, usually implemented by means of inter governmental agreements, plays a large role in the development and the operation of space hardware, especially in very large facilities such as the International Space Station (ISS). Coordination activities, like e.g. the International Exploration Coordination Group\protect\footnote{\protect\url{https://www.globalspaceexploration.org/wordpress/}} (ISECG) are also crucial for the implementation of future highly ambitious space projects, such as human and robotic exploration of the solar system, and the development of bases on the Moon, and Mars. These initiatives clearly hold a large technical, political and economic value.

There are no similarly large and structured examples of international cooperation projects in space science data management, processing and archiving. However, especially in the astronomy sector, some important examples of cooperation emerged in the form of spontaneous aggregation of international research institutions and individuals. Perhaps the most notable one is the cooperation that led to the development of the FITS (Flexible Image Transport System) format. This data format was originally designed in the 1970’s for the exchange of radio telescopes images, and over the years, and through the collaboration of experts from space and ground based telescopes operating in all energy bands, evolved into a standard capable of supporting any type of astronomical data type, and endorsed by all major organizations, e.g. NASA, ESA and The International Astronomical Union (IAU).

Today the FITS format is not only a consolidated asset for space science and astronomy, but also a valuable resource that can be used for other purposes. One important example is the use of FITS\protect\footnote{See the NASA-maintained FITS support office at \protect\url{https://fits.gsfc.nasa.gov}}) for the preservation of high-resolution images of ancient documents to ensure that present and future generations will have simple access to the books of the Vatican library\protect\footnote{A description of the Vatican Library Long Time Digital Preservation project can be found here: \protect\url{https://www.vaticanlibrary.va/home.php?pag=digitalizzazione&amp;ling=eng}}.

Another important example of cooperation is the International Virtual Observatory Alliance (IVOA\protect\footnote{See \protect\url{http://ivoa.net}}), which, since 2002, focuses on the development of standards and services to the astronomical community to support good data management and inter-operability. The IVOA currently comprises 21 programmes from Argentina, Armenia, Australia, Brazil, Canada, Chile, China, Europe, France, Germany, Hungary, India, Italy, Japan, Russia, South Africa, Spain, Ukraine, the United Kingdom, and the United States, and an inter-governmental organization -- ESA.

Open Universe, with the coordination of UNOOSA, wishes to largely expand the current level of international cooperation in space science data services, with large benefits for the scientific community, and other sectors of the society. This will be achieved by aggregating another level of international cooperation and coordination, centred at the United Nations and its broad network of member states and international organisations.

\section{Ongoing Open Universe Activities}\label{actvities}
Following the Vienna workshop in November 2017, three main lines of activities have been pursued in an effort to implement the vision of the initiative: (i) The consolidation of the prototype Open Universe portal\protect\footnote{\protect\url{https://openuniverse.asi.it}}, developed at the Italian Space Agency; (ii) The set up of a prototype implementation of a technical infrastructure in a number of participating countries and institutes (e.g., Armenia-Yerevan, Argentina-La Plata, Brazil-CBPF, IcraNet-Pescara, Italy-ASI, NYU-Abu Dhabi) consisting of hardware and services that will eventually constitute the Open Universe operational nodes, with an aimed focal point at UNOOSA; and (iii) Scientific and educational activities aimed at demonstrating the feasibility of the principles and some of the long-term objectives of the initiative, both in terms of technical implementation and of cost sustainability. All these activities are conducted in the spirit of a proof-of-concept and establishment of the international collaboration that will serve as seed for the Initiative.

Most of the current work is carried out by the team of researchers participating to Open Universe on a best-effort basis, and largely within their research time and with resources from their regular research funds.

\subsection{The ASI Open Universe portal}
As a first step  towards  the  implementation  of  the  principles  put forward by ”Open Universe”, the Italian Space Agency (ASI) developed a prototype web portal for the initiative\protect\footnote{The portal is supported by ancillary distributed infrastructure such as provided by the Brazilian Science Data Center (BSDC) hosted at CBPF, in Rio de Janeiro -- see \protect\url{http://vo.bsdc.icranet.org}.}. The first version of the ASI ”Open Universe” portal was released on the occasion of the United Nations/Italy workshop on the Open Universe held in Vienna in November 2017.
The main aims of this service are:
\begin{itemize}
    \item  develop the prototype that concentrates, in a single web page, the potential of accessing space science data and information in a transparent way from several data archives and information systems
    \item provide links to a large number of services that give access to space science data resources.
    \item facilitate  new  types  of  scientific  research  based  on  data  intensive/data  fusion analysis
    \item  help defining the requirements for a new generation of ”user-centred” integrated space-science data archives that could be used by anyone with access to mobile technologies.
    \item expose the potential of ”Open Universe” to the non-space science professionals (e.g. museums, education sector, citizen citizens).
    \item provide access to Open Universe developed tools that simplify the analysis and use of space science and astronomy data 
\end{itemize}

The portal, available at the following URL https://openuniverse.asi.it, 
has been conceived as a sort of space science data ”shopping mall” where users visit and use (shop) the many web services (each clearly identified by its ”brand” or developers' logos) available next to each other, with the peculiarity that all the services (shops) know what the user is looking for as soon as he/she enters the portal and specifies a source name or a position in the sky.

\subsection{Space science for non-specialists, non-experts and the citizen}
In an effort to remove, or at least significantly lower some of the existing barriers to space science data, and therefore increase data transparency, we have been developing software tools that aim at encapsulating complexity and hiding it from the users.

We have done so using modern and emerging technologies, like e.g. Linux containers, in particular from Docker.com\protect\footnote{\protect\url{https://www.docker.com}}, and the services and tools provided by the International Virtual Observatory Alliance  (IVOA)\protect\footnote{\protect\url{http://www.ivoa.net}}.

\begin{figure}[th!]
  \includegraphics[angle=-90, width=\textwidth]{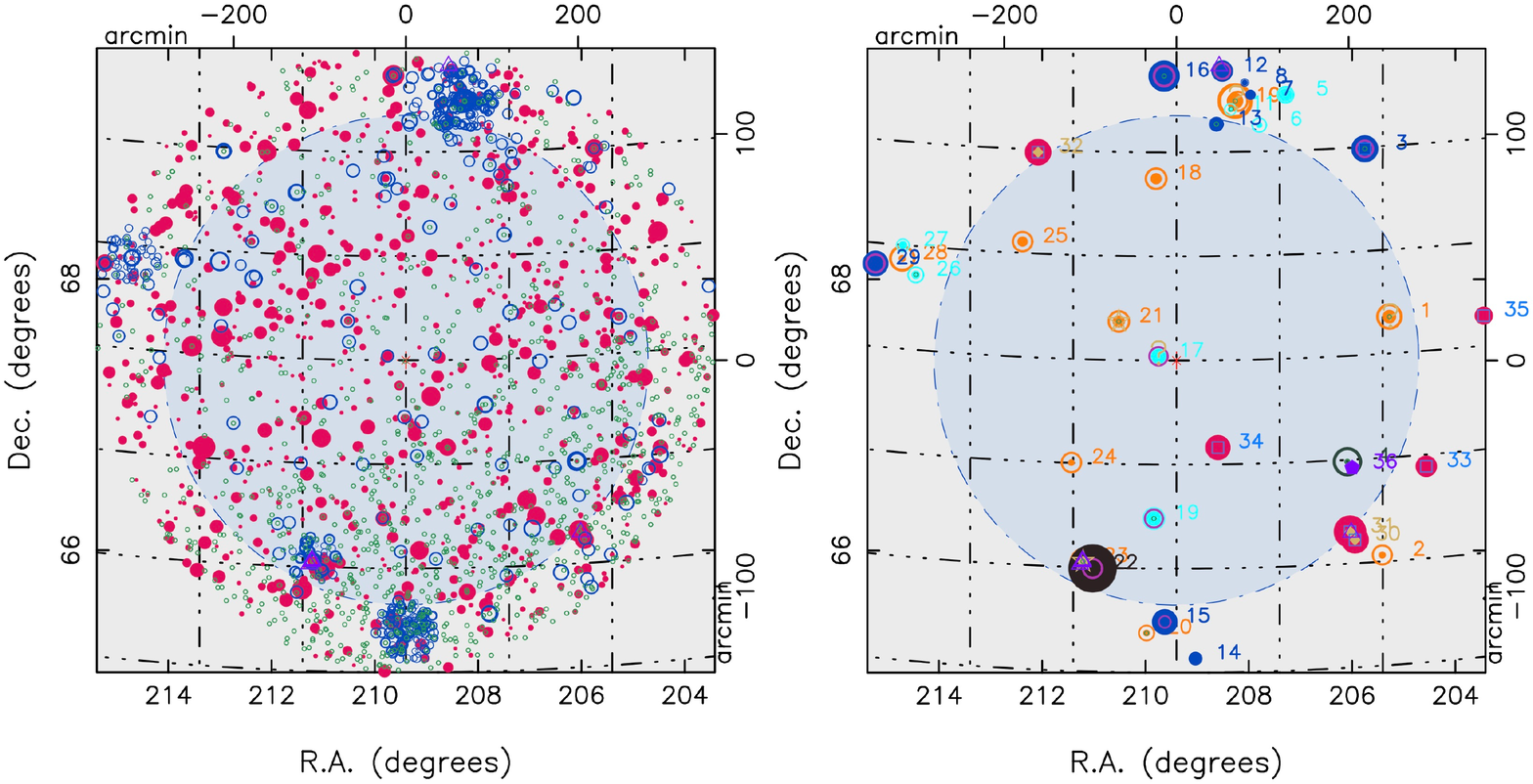}
  \centering
  \caption{The region around the uncertainty in the arrival direction of the IceCube neutrino IC140216A (light blue elliptical area) as presented by VOU-Blazars. Left: multi-frequency catalogued sources, right: known and candidate blazars. See text for details).}
  \centering
  \label{Fig:maps}
\end{figure}

\begin{figure}[th!]
  \includegraphics[angle=-90, width=\textwidth]{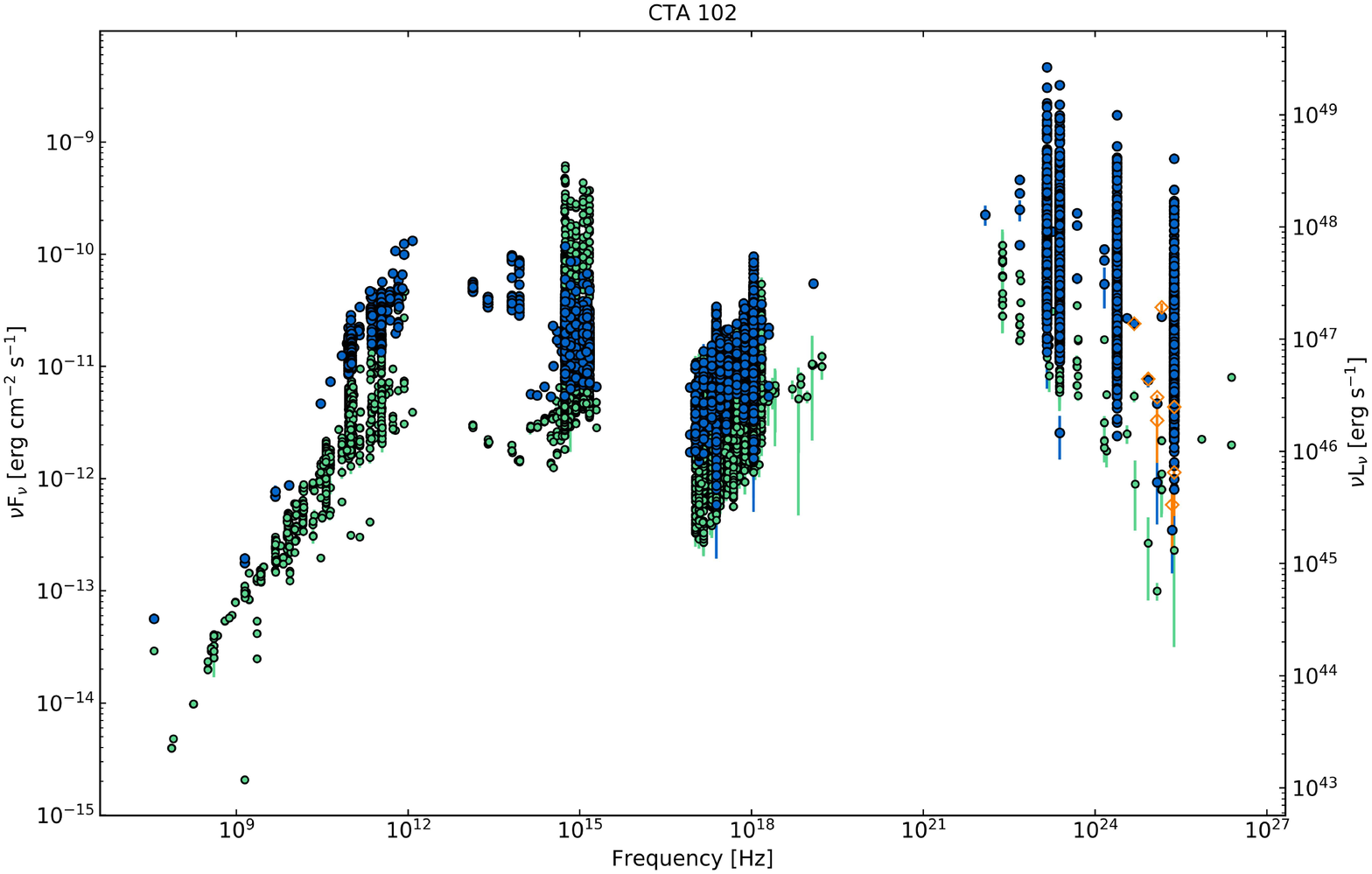}
  \centering
  \caption{Radio to gamma-ray Spectral Energy Distribution of the blazar CTA102, one of the most luminous sources in the sky. Note the very large variability present at far-infrared, optical, X-ray and $\gamma$-ray frequencies.}
  \centering
  \label{Fig:SED}
\end{figure}

Among the tools developed so far we mention: 

- {\bf VOU-Blazars} \cite{Chang2020}, is a software tool that, using Virtual Observatory (VO) services, retrieves data from dozens of  catalogs of astronomical sources detected at all wavelengths and, based on the information retrieved, builds images of the sky highlighting the position of possible blazars.
This is particular useful to visualize or discover new blazars, especially in the large or very large error regions of $\gamma$-ray sources or high-energy neutrino events.
As an example Fig. \ref{Fig:maps} shows the area covering the uncertainty region of the IceCube neutrino IC140216A. The left image plots the radio (red filled circles), X-ray (open blue circles), $\gamma$-ray (open purple triangles) and known AGN (small green circles). The right image plots the multi-frequency matching sources that could be blazars. Different colours represent possible blazars of different SED type.
For each candidate blazar the tool can build a detailed radio to $\gamma$-ray Spectral Energy Distribution (SED) (See \cite{Chang2020} for details). The spectral and time-domain data obtained with VOU-Blazars can be easily combined with data from other on-line tools. Fig \ref{Fig:SED} is an example of SED of the bright blazar CTA102 built with VOU-Blazar data (blue points) and from other tools (green points).

- {\bf Swift\_deepsky} \cite{Giommi2019} is  a  software  pipeline  built  on  top  of  the Swift X-Ray Telescope (XRT) official Data Analysis Software (XRTDAS), that  performs  a  complete  X-ray image analysis of Swift XRT data. 
The software is publicly available either as open source code or as encapsulated in a Docker container ready to be used on most platforms, including MacOS, Linux and Windows10. Details on how to run the software and the products generated can be found in \cite{Giommi2019}.
Swift\_deepsky has been run on various large samples of Swift observations and the results have been made public,
including the all observations of blazars \cite{Giommi2019}, a sample of over 1,300 Gamma Ray Bursts \cite{Giommi2020} and for the generation of stacked X-ray images of the full data set (15.5 years) of Swift-XRT observations in photon counting (imaging) mode. This last
run is very recent and took about 6 days of processing times in the currently existing six Open Universe proto-nodes.

- {\bf Swift\_xrtproc} \cite{Giommi2015},\cite{Giommi2020b}, is a software tool 
based on the official Swift Software that performs a detailed spectral analysis of X-ray sources. The tool is currently being extensively tested and prepared for a Docker container version to be released in the coming months. 

To increase the level of transparency of open access data, making them more easily available to the non-specialist, we ran these tools in an automatic way on large parts of the Swift archive, simulating scientific work carried out by non-experts. This resulted in new, ready to use blazars imaging and broad-band spectral data \cite{Giommi2019}, detailed spectral and time domain results of the sample of blazars that have been observed with XRT more than 50 times \cite{Giommi2020b}, a new medium-deep survey of the X-ray sky, and a complete deep sample of HBL sources \cite{Giommi2020}, the rarest type of blazars.

Similar projects, based on archival data from the hard-X-ray and $\gamma$-ray telescopes NuSTAR and Fermi-LAT are on-going and the results will be presented in the near future.

\subsection{Education: Open Universe workshops/schools}
One of the main objectives of Open Universe is to "Broaden the user base", that is, significantly widen the community of users of space science data. That long-term goal is pursued from different perspectives. One important activity on this front is the organization of workshops and schools. As of today two such events have already taken place. 

\begin{itemize}
    \item The Open Universe International Doctoral School: "The discovery of Black Holes" at the ICRANet Centre in Nice. Mostly dedicated to PhD students, postdocs and scholars\protect\footnote{\protect\url{http://www.icranet.org/index.php?option=com\_content\&task=view\&id=1241}}.
    \item Open Universe workshop at New York University - Abu Dhabi
    Dedicated to undergraduate students\protect\footnote{https://sites.google.com/view/ouworkshop-nyuad/home}.
\end{itemize}

In both cases the innovative tools and data access methods so far 
developed within the initiative or available through the interface of the open universe portal were described and demonstrated. 

\section{Sustainable Costs}\label{sustainability}
We argue that initiatives such as "Open Universe" can contribute to contain the costs associated to large international cooperative enterprises, such as the BRICS Astronomy Programme on Transients\protect\footnote{See the Resolutions of the Fourth Meeting of the BRICS Astronomy Working Group: \protect\url{https://www.bricsastronomy.org/wp-content/uploads/2018/11/Final-Minute-and-Resolutions-of-the-4th-BRICS-Astronomy-Working-Group-Me....pdf}}, making them more sustainable and effective.

In the long run, when considering future facilities, this can be achieved by ensuring that the final data products meet openness and transparency requirements, which in turn boost science productivity, facilitate international cooperation, and increase socioeconomic and educational impact of the activities.

In addition to that, when considering, upstart, the integration of existing facilities and establishing new collaboration avenues among the BRICS countries, cost-efficiency can be ensured through the Open Universe by:

\itemize{
\item Avoiding duplication of efforts amongst partners in different countries
\item Fostering collaboration and coordination among existing data centres and observatories
\item Making best use of existing high quality infrastructure and data services (e.g. full use of IVOA standards and tools whenever possible)
\item Facilitating the development of innovative tools, and of a new paradigm, such as distributed analysis 
\item Providing the appropriate background for taking full benefit of new technologies\\
}

\section{Outlook on BRICS Cooperation and Role in the Open Universe}\label{BRICS}
As emerging economies, the BRICS countries face a specific set of challenges, but are also uniquely placed to act on the opportunities that the fourth industrial revolution presents. The Open Universe is an Initiative framed around the development imperatives of the United Nations Sustainable Development Agenda -- targeting particular direct impact in Quality Education (SDG4) and the strengthening of International Partnerships (SDG17). The BRICS countries are well placed to take a global lead in the conduction of the Initiative, and its associated development agenda, within the United Nations, with great benefit for the development of space sciences and scientific cooperation in the Bloc.

As the Initiative is part of the Capacity Building Framework of UNOOSA, the BRICS countries can further benefit from Open Universe's strong component of capacity building actions for the 4th Industrial Revolution, through the training of students and young researchers, seeking to use space sciences to address in-country socioeconomic development challenges.

At the core of BAWG activities is the development of a network of astronomical research among the BRICS countries, for which Open Universe can provide part of the data infrastructure necessary to enable the BRICS Astronomy science and development programmes. In particular, it can directly benefit from the support of the Open Universe Initiative for public release and access to data. Given the focus of the BRICS Astronomy Flagship Project in survey and transients sciences, from one side, and big data management on the other, the technical focus of the Open Universe Initiative on improving transparency, integrability, and accessibility to astronomical databases, has a clearly enabling role both in the scientific and the education/outreach fronts of the BRICS Astronomy research programme.

\section{Summary and conclusions}\label{conclusions}

Since the advent of the first web-based digital archives offering on-line open astronomical data services in the early nineties, much has been done in the direction of offering space science data to an ever increasing number of users, from the small community of scientists involved in the experiments that produced the data, to several thousands of non-specialists researchers. This progress, however, has been strongly discipline-dependent, with the astronomy sector leading the way, while other space science disciplines moving at a much lower speed, and in many cases still restricting the use of the expensive data they produce to the small number of scientists belonging to the project teams.

The Open Universe initiative aims to play a strategic role in this regard in the near future, by improving the use of digital technologies to maximize transparency and enhance web interfaces with applications that may even go beyond the field of space science and astronomy. Today’s best digital archives provide, in most cases, calibrated data and the associated software suitable for higher level scientific analysis that must be carried out by expert users. The rest of the world is still largely excluded from the utilisation of the data, particularly in emerging and developing countries such as is the case of the BRICS. It is part of our goals to contribute to reverse this picture.

From the  point of view of its development, 2016 was the year of the proposal to COPUOS, while 2017 was dedicated to ample discussions, with the organization of an expert meeting in Rome \cite{ASI2017}\protect\footnote{See the presentations of the ASI Open Universe Expert Meeting, available here: \protect\url{https://openuniverse.asi.it/documents/ou\_documents.php}}, and a workshop dedicated to Open Universe open to the international community that took place at the United Nations in Vienna \cite{UNOOSA2017}\protect\footnote{See the presentations of the UN-Italy Workshop on the Open Universe Initiative here:\protect\url{unoosa.org/oosa/en/ourwork/psa/schedule/2017/workshop\_openuniverse\_presentations.html}}. A second expert meeting is planned for the end of the year in Brazil, with significant contribution from the BRICS countries.

In 2020 Open Universe will undergo formal foundation within UNOOSA, and enter its operational phase as part of the activities of the United Nations Space 2030 Agenda. It is aimed that the Initiative will closely cooperate with the BRICS Astronomy Working Group, as well as scientists and institutions at large from the BRICS countries, to provide services and support the development of the BRICS Astronomy Programme and its foreseen educational and social activities.

\section{Acknowledgments}\label{acknowledgments}
UBdA acknowledges the support of a Serrapilheira Institute Grant number Serra - 1812-26906.\\
PG acknowledges support of the Technische Universit{\"a}t M{\"u}nchen - Institute for Advanced Study, funded by the German
Excellence Initiative (and the European Union Seventh Framework Programme under grant agreement n. 291763).


\end{document}